\documentclass[a4paper]{isp}
\usepackage{graphicx}
\def\aap{A\&A\,  }
\def\apj{ApJ\,  }

\def\mnras{MNRAS\,  }

\def\physa{Phys. A    }
 
\def\za{Z. Astrophys.  } 
\def\h0units{\mathrm{km\,s^{-1}\,Mpc^{-1}}}
\def\cunits{\mathrm{km\,s^{-1}}}

\newcommand{\dl}{d_{L}}
\newcommand{\pl}{d_{L,2,2}}
\newcommand{\ml}{d_{L,m,2,2}}
\newcommand{\om}{\Omega_{\rm M}}

\begin{document}

\title
{
Voronoi chords in flat cosmology
}
\author{Lorenzo Zaninetti}

\institute
{
Physics Department  ,
 via P.Giuria 1,\\ I-10125 Turin,Italy 
}

\maketitle

\begin{abstract}
In this paper we present formulae 
for  chord length  distribution 
in the framework  of Poissonian Voronoi Tessellation (PVT)
and non Poissonian Voronoi Tessellation (NPVT).
The introduction of the scale parameter in the obtained
distributions 
allows us to model the chord for cosmic voids.
A graphical comparison  between 
cosmic voids visible  on two catalogs of galaxies,
2dFGRS and VIPERS, and theoretical random chords
is reported.
\keywords
{  
Cosmology;
Observational cosmology;
Distances, redshifts, radial velocities, spatial distribution of
galaxies;
}
\end{abstract}

\section{Introduction}

A first  study  on  the random 3D fragmentation 
by \cite{Kiang1966}
was done  in
the framework of  Voronoi Diagrams.
In the above paper  the Kiang's conjecture 
was introduced: 
"The normalized probability density function for volumes
in Poissonian Voronoi Tessellation 
follows a normalized gamma distribution 
with shape parameter $c$ equal to six".
A subsequent 
 large-scale computer simulation 
 on Poissonian Voronoi Tessellation (PVT),
 see \cite{Ferenc_2007}, fixed in $c=5$ the shape parameter of 
the normalized gamma distribution, in the following
Kiang's function
The shape  parameter $c$ plays 
an important role in the determination 
of the probability density function (PDF) 
which model the Poissonian and non Poissonian Voronoi Tessellation
(NPVT) chord,  see \cite{Zaninetti2013b,Zaninetti2015b}.
The 3D tessellation  is built 
in the framework of the Euclidean distance 
conversely in astronomy  the observable parameter is
the redshift $z$.
In cosmology a key target is to deduce
the luminosity distance as a function of the redshift,
see as an example \cite{Zaninetti2016a}
where different models were analyzed.
The inverse problem,  i.e.,  
the redshift as function of the luminosity 
distance,  is poorly known and this
fact has stopped the application 
of the Voronoi Diagrams to cosmology.
The enormous progresses in observational cosmology
allows to model the radius of cosmic voids, see
\cite{Vogeley2012,Varela2012}.
A statistical analysis of these catalogs allows the
determination of the shape parameter $c$ of the Kiang's function.
This paper 
explores in Section \ref{secbasic}
the fundamentals of chord for spheres and reviews 
the existing knowledge for  volumes in PVT and NPVT.
In section \ref{secchords}
a PDF  for chord in PVT and NPVT is investigated.   
The luminosity distance as function of the redshift
in flat cosmology  and the corresponding 
inverse function is derived in Section 
\ref{secflatcosmo}.
Section \ref{secastrophysical}
reports a comparison between real voids of galaxies
and theoretical random cosmic chord.

\section{The basic formulae}
\label{secbasic}

This Section reviews the basic formulae for 
chords of spheres  and  the distribution in volumes 
for  PVT and NPVT.

\subsection{The chord}

The average length, $<l>$,
of all chords
of spheres  having the same radius $R$
\begin{equation}
<l> = \frac{4}{3} R
\label{monogeometrical}
\quad  ,
\end{equation}
see more details in \cite{Zaninetti2013b}.

Given  a probability density function
(PDF)  for the diameter of the voids,
$F(x)$,  where   $x$  indicates  the diameter.
The  probability, $G(x)dx$,
that  a sphere having diameter between
$x$ and  $x+dx$ intersects  a random line is
proportional  to their  cross section
\begin{equation}
G(x) dx  =  \frac { \frac{\pi}{4} x^2  F(x) dx }
                  { \int_0 ^{\infty}
                    \frac{\pi}{4} x^2  F(x) dx}
=
\frac {  x^2  F(x) dx }
      {  < x^2>}
\quad  .
\end{equation}
Given a line  which intersects a sphere of diameter
$x$, the probability that the distance
from the center  lies  in the range
$r,r+dr$  is
\begin{equation}
p(r) = \frac{2 \pi r dr }{\frac{\pi}{4} x^2 }
\quad ,
\end{equation}
and the chord length is
\begin{equation}
l = \sqrt { x^2 - 4r^2}
\quad .
\end{equation}
The probability that  spheres  in the
range  $(x,x+dx)$   are intersected to produce
chords  with lengths  in the range
$(l,l+dl)$ is
\begin{equation}
G(x)\, dx  \frac{2l\,dl}{x^2}
=
\frac{2l \, dl} { <x^2>}  F(x) dx
\quad .
\end{equation}
The probability  of having a chord
with  length  between $(l,l+dl)$  is
\begin{equation}
g(l)
=
\frac{2l} { <x^2>}  \int_l^{\infty} F(x) dx
\quad .
\label{fundamental}
\end{equation}
This integral  will be called {\it fundamental}
and the previous  
demonstration  follows \cite{Ruan1988,Zaninetti2013b}.

\subsection{Voronoi Diagrams}

\begin{figure}
\begin{center}
\includegraphics[width=10cm]{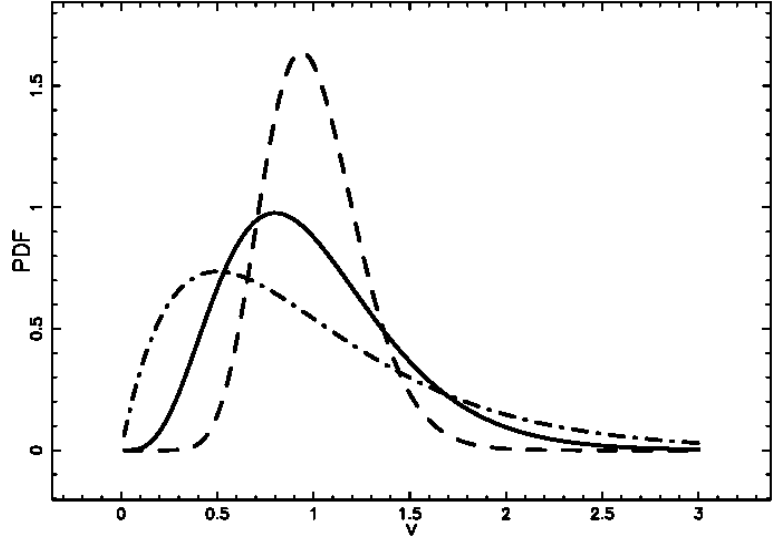}
\end{center}
\caption
{
Reduced  volume distribution when c=5  (full line),
c=16 (dashed line) and c=2 (dot-dash-dot-dash line).
}
\label{volumi_c}
\end{figure}

The Voronoi
tessellation is the partition of space
for a given  seeds  pattern
and the result of the partition depends completely on
the type of given pattern
"random",
 Poisson-Voronoi tessellations
 (PVT),
   or "non-random",
non Poisson-Voronoi tessellations
 (NPVT).
 The reduced volumes of Voronoi cells generated with
 random seeds
 can be fitted by
the Kiang function
\begin{equation}
 p(x;c) = \frac {c^c}{\Gamma (c)}
x^{c-1}
\exp(-cx),
\label{kiang}
\end{equation}
see
\cite{Kiang1966}, which has variance
\begin{equation}
\sigma^2 =\frac{1}{c}.
 \end{equation}
It has shown that good approximations for volume
distributions of
PVT  cells can be obtained by setting
$c=5$  \cite{Ferenc_2007,Zaninetti2015b}.
The case of more regular partition of the
space is produced by
the Sobol seeds which produce  a distribution
in volumes with $c \approx 16$.
Conversely the introduction of seeds generated
on a adjustable  Cartesian grid  (ACG) allows
to cover the interval  $c \in (2-16)$ in a continuous
way,  see \cite{Zaninetti2015b}.
Figure \ref{volumi_c} reports
three  cases of volumes for the Voronoi Diagrams
from which is clear the transition from 
order to disorder as function of decreasing c.

\section{The generalized chord}

\label{secchords}
We start by approximating the reduced volumes
of the Voronoi tessellation
by a Kiang function,
therefore the   distribution in diameters,  $y$,  is
\begin{equation}
F(y)=
\frac
{
c \left( \frac{1}{6}\,c\pi\,{{\it y}}^{3} \right) ^{c-1}{{\rm e}^{-\frac{1}{6}\,c\pi
\,{{\it y}}^{3}}}\pi\,{{\it y}}^{2}
}
{
2\,\Gamma \left( c \right)
}
\quad  ,
\end{equation}
where
\begin{equation}
\mathop{\Gamma\/}\nolimits\!\left(x\right)
=\int_{0}^{\infty}e^{{-t}}t^{{x-1}}dt
\quad ,
\end{equation}
is the gamma function.
We  insert in the fundamental
Equation (\ref{fundamental}) for the chord  length
the  generalized PDF for the diameters.
The resulting integral  is
\begin{eqnarray}
g(l;c)=
\frac
{
-{\pi}^{2/3}\sqrt [3]{6}l \left( {6}^{-c}{c}^{c}{\pi}^{c}{l}^{3\,c}{
{\rm e}^{-1/6\,c\pi\,{l}^{3}}}-\Gamma \left( 1+c,1/6\,c\pi\,{l}^{3}
 \right)  \right)
}
{
3\,\sqrt [3]{c}\Gamma \left( 2/3+c \right)
}
\quad ,
\end{eqnarray}
where
\begin{equation}
\mathop{\Gamma\/}\nolimits\!\left(a,z\right)=\int_{z}^{\infty}t^{{a-1}}e^{{-t}%
}dt,
\end{equation}
is the  upper incomplete gamma function,
see
\cite{Abramowitz1965,NIST2010}.

\begin{figure}
\begin{center}
\includegraphics[width=10cm]{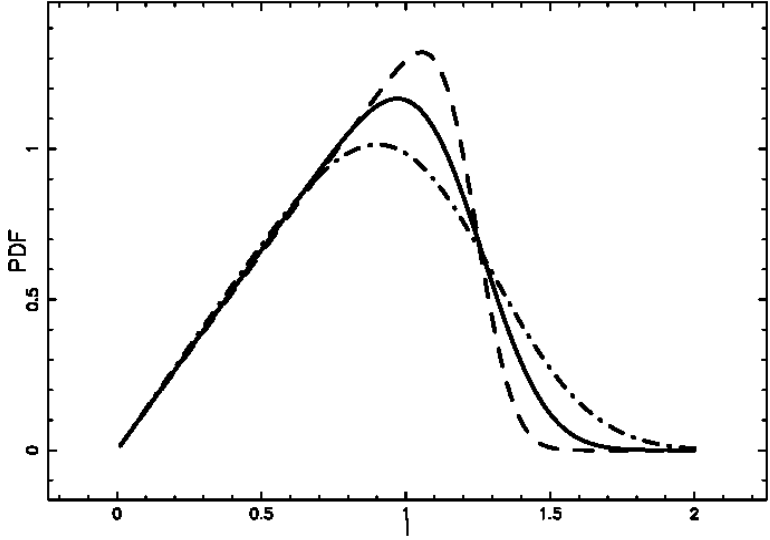}
\end{center}
\caption
{
Chord PDF for Voronoi Diagrams, g(l;c), when  c=5  (full line),
c=16 (dashed line) and c=2 (dot-dash-dot-dash line)
}
\label{gl_b_cvar}
\end{figure}

\begin{figure}
\begin{center}
\includegraphics[width=10cm]{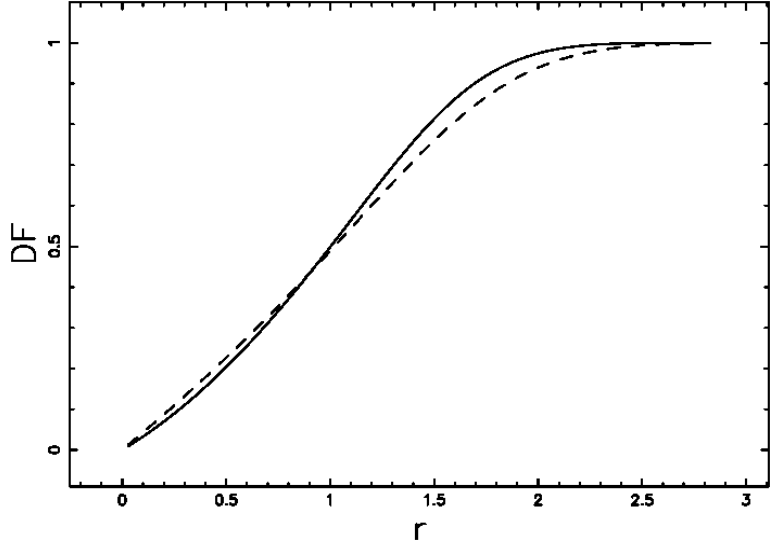}
\end{center}
\caption
{
The full line represents the tabulated
chord  DF  in the case of PVT,
the dashed line is our
chord approximate  $DF_5$.
}
\label{dfpvtmuche}
\end{figure}

Figure \ref{gl_b_cvar} reports
three  cases of $g(l;c)$  as function of  $c$.
We have approximated
the volumes of the Voronoi diagrams by spheres
and  the length
of the chord which touches only in one point
the sphere is zero,  conversely
a chord which lies on a irregular face of
the Voronoi's polyhedron
has a finite length.
In  order to solve this inconvenient a shift
should  be introduced.
A new  variable  $z$ has
been defined by a shift of $l$: $z=l-a$, so that
\begin{equation}
g_s(z;c,a)=g(z+a;c,a)
\quad ,
\end{equation}
where $a$ is  the shift parameter
and $g_s(z;c,a)$ is the shifted PDF for Voronoi's chord.
Next, to obtain a reduced variable,
average value equal one or an astrophysical
variable, a scale change has been
applied ,$u =bz$,
resulting in scaled PDF for chords 
\begin{equation}
g_b(u;c,a,b)= \frac{g_s(\frac{u} {b};c,a)} {b}
\label{gbscaled}
\quad .
\end{equation}
As an example the PVT PDF for chords, $c=5$, can be
obtained inserting $a=0.4068$ and $b$ =1.9626
and is
\begin{eqnarray}
g_b(u;5,0.4068,1.9626)=
  (  0.3690+ 0.0002321\,{u}^{13}+ 0.002410\,{u}^{12}+ 0.01155\,{u}
^{11}+ 0.03649\,{u}^{10}
\nonumber \\
+ 0.08890\,{u}^{9}+ 0.1739\,{u}^{8}+ 0.2903\,{
u}^{7}+ 0.4413\,{u}^{6}+ 0.5798\,{u}^{5}+ 0.7160\,{u}^{4}
\nonumber \\
+ 0.8327\,{u}
^{3}+ 0.6931\,{u}^{2}+ 0.7065\,u   ) {{\rm e}^{- 0.3463\,   ( u
+ 0.7985   ) ^{3}}}
\quad  .
\end{eqnarray}
The above PDF in the interval $u \in (0,3.2)$ is normalized to
one and has average value one.
The distribution function (DF) of this  PDF does not exists 
but the best minimax rational approximation of degree (5, 2)
allows to find the following approximate distribution 
function, $DF_5$,
\begin{equation}
DF_5(x) =
\frac{  0.4446+ 0.1266\,x- 0.4449\,{x}^{2}+ 0.1910\,{x}^{3}- 0.02560\,{x}^{4}
+ 0.0004040\,{x}^{5}
}{ 1.447- 1.462\,x+ 0.4596\,{x}^{2}}
\quad .
\end{equation}
Figure   \ref{dfpvtmuche}
reports a comparison of the previous
approximate DF,
as a dashed line
with the tabulated result
as
deduced from Table 5.7.4  in \cite{Okabe2000}
when the average value of both  PDFs is one.

\section{Flat cosmologies with cosmological constant}
\label{secflatcosmo}

As eqn.(2.1) in \cite{Adachi2012}
the luminosity distance $dl$ is
\begin{equation}
  \dl(z;c,H_0,\om) = \frac{c}{H_0} (1+z) \int_{\frac{1}{1+z}}^1
  \frac{da}{\sqrt{\om a + (1-\om) a^4}} \quad ,
  \label{lumdistflat}
\end{equation}
where $H_0$
is the Hubble constant expressed in     $\h0units$,
$c$ is the light velocity  expressed in $\cunits$,
$z$ is the redshift,
$a$ is the scale-factor
and  $\om$ is
\begin{equation}
\om = \frac{8\pi\,G\,\rho_0}{3\,H_0^2}
\quad ,
\end{equation}
where $G$ is the Newtonian gravitational constant and
$\rho_0$ is the mass density at the present time.
The above integral does not have an analytical simple 
formula but  the Pad\'e approximation of degree $m,n$ 
about the point $a$=1 of the integrand allows to 
solve  in an  approximate way the integral.

We report the Pad\'e approximate integral
of the luminosity distance 
$\pl$, in the case of  $m$=2 and $n$=2 when 
$H_0 = 100 \h0units$ and $\om=0.3$ as  in 
\cite{Varela2012} 
\begin{eqnarray}
\pl(z)   
=
\Re \bigg (
2997.924\,   ( 1+z   )    ( - 0.0953787\,   ( 1+z
   ) ^{-1}+ 1.04035- 0.55174\,i+
\nonumber \\
 0.175624\,\ln    ( -
 624.128\,   ( 1+z   ) ^{-2}+ 92.2245\,   ( 1+z
   ) ^{-1}- 284.222   ) 
\nonumber \\
- 2.248466\,\arctan   ( 
 1.490823\,   ( 1+z   ) ^{-1}- 0.110146   ) 
   )
\bigg ) 
\quad . 
\label{dlpade22}
\end{eqnarray}
In the above  complex analytical solution we have   a real part 
denoted by $\Re$ 
and a negligible imaginary part.
As an example the above real part is 
23977.70565   when $z=4$ 
and the imaginary  part is $0.5546 \, 10^{-4} $.

In the above formula $\pl$ is function,  $f$,  of $z$ 
\begin{equation}
\pl= f(z)
\quad  .
\end{equation} 
The inverse function $f^{-1}$ is 
\begin{equation}
z = f^{-1} (\pl) 
\end{equation}
but eqn. (\ref{dlpade22}) is not invertible for $z$.

In order to have an inverse function we apply the
best minimax rational approximation of degree (m, n) 
to eqn.(\ref{dlpade22}) over the interval $z \, \in (0,4)$.
In the case of $m$=2 and $n$=2 the minimax rational expression
for the luminosity distance, $\ml$, 
is  
\begin{equation}
\ml =
\frac{
1.792+ 1292.747\,z+ 2020.755\,{z}^{2}
}
{
0.4463702+ 0.2686967\,z+ 0.002703757\,{z}^{2}
}
\quad .
\label{dlfzminimax}
\end{equation}
The inverse function exists and is
\begin{equation}
z=\frac{N(d_L)}{D(d_L)}
\label{zinverse}
\quad ,
\end{equation}
where
\begin{eqnarray}
N=
- 6.71741\,10^9\,{\it d_L}+ 3.231869\,10^{13}
\nonumber\\
-\sqrt {{
 4.210652\times 10^{19}}\,{{\it d_L}}^{2}+{ 1.820828\times 
10^{24}}\,{\it d_L}+{ 1.035444754\times 10^{27}}}
\nonumber
\end{eqnarray}
and
\begin{equation}
D=
1.35187\,10^8 \,{\it d_L}- 1.010377933\,10^{14}
\quad .
\nonumber
\end{equation}

In the framework of a numerical approximation for the inverse function 
\begin{equation}
f(f^{-1}(x_1)) = x2
\quad ,
\end{equation}
with $x_1 \approx x_2$,
we  introduce the percentage error, $\delta$, of the inversion
formula (\ref{zinverse}) 
\begin{equation}
\delta = \frac
{
\big | x_1 - x_2 \big |
}
{
X_1
} \,  100 
\quad .
\end{equation}
In our case when $x_1=1200\, Mpc$
$f(f^{-1}(1200\,Mpc)) = 1203.4\, Mpc$ which means 
$\delta = 0.027 \%$.

\section{Astrophysical Applications}
\label{secastrophysical}

This section  processes  two astronomical catalogs  for cosmic 
voids, deduces  PDF and DF for astrophysical chord,
introduces the cosmic chord and
randomly generates  chords on two catalogs of
galaxies.

\subsection{The catalogs of voids}

A {\it first} catalog of cosmic voids  
can be found 
in \cite{Vogeley2012}
where the effective radius of the voids has been derived.
The main results, the averaged radius of the voids 
and the value 
of $c$ for the reduced volumes of spheres,
are
\begin{equation}
R = 18.23 h^{-1}\, Mpc \quad ;\quad c=1.458  \quad Pam\, et \, al. \,2012
\quad ,
\end{equation}
we call it the non Poissonian case.  
A {\it second} catalog is that of  radii larger than 10h-1Mpc 
up to redshift
$0.12\,h^{-1} Mpc$   in (SDSS-DR7), see \cite{Varela2012}.
In this case  the average radius and $c$ are 
\begin{equation}
R = 11.85 h^{-1}\, Mpc \quad ;\quad c=4.763  \quad Varela\, et \, al. \,2012
\quad ,
\end{equation}  
we call it the nearly  Poissonian case, 
being the Poissonian case $c$=5.

\subsection{Astrophysical chords}

An astrophysical version for the PDF of chords can be obtained 
inserting in eqn.\ref{gbscaled}
the appropriate scale, $b$, and the shape parameters, $c$, 
see Table \ref{bcparameters}  
\begin{table}[ht!]
\caption
{
Parameters
of the astrophysical chords.
}
\label{bcparameters}
\begin{center}
{
\begin{tabular}{|c|c|c|c|c|}
\hline
Case                & a       &  b (Mpc)   & c     & $<l>$ (Mpc)\\
Nearly~ Poissonian  & .402    &  30.7664   & 4.763 & 15.799   \\
Non~Poissonian      & .345    &  40.2374   & 1.458 & 24.306   \\
\hline
\end{tabular}
}
\end{center}
\end{table}

The PDF which models  the chord for  cosmic voids in the 
nearly Poissonian case is 
\begin{eqnarray}
g_{b,nearly}(x)=  
- 0.00033275\,\sqrt [3]{6}   (  0.032502\,x+ 0.40268
   )  \bigg (  77.24\,   (  0.032502\,x+ 0.40268
   ) ^{ 14.28}
\nonumber \\
{{\rm e}^{- 2.49233\,   (  0.0325\,x+
 0.40268   ) ^{3}}}-\Gamma   (  5.76, 2.49233\,
   (  0.032502\,x+ 0.40268   ) ^{3}   )  \bigg ) 
\quad nearly ~ Poissonian
\quad ,
\label{pdfastrophysicalnearly}
\end{eqnarray}
and in the non Poissonian case is 
\begin{eqnarray}
g_{b,non}(x)=
-0.0162275\,\sqrt [3]{6}   (  0.0248524\,x+ 0.345018
   )  \bigg   (  0.670705\,   (  0.0248524\,x+
 0.345018   ) ^{ 4.35}
\nonumber \\
{{\rm e}^{- 0.759218\,   ( 
 0.0248524\,x+ 0.345   ) ^{3}}}-\Gamma   (  2.45,
 0.759218\,   (  0.024852\,x+ 0.345   ) ^{3}
   )  \bigg  ) 
\quad non ~Poissonian 
\quad .
\label{pdfastrophysicalnon}
\end{eqnarray}
The two previous PDFs cannot be integrated 
and therefore the 
analytical DFs do  not exist.
We therefore {\it first} deduce two approximate PDFs 
trough the best minimax rational approximation of degree (5, 2).
The integration of the two approximate PDFs gives 
\begin{eqnarray}
DF_{nearly}(x) =
- 0.0000036854\,{x}^{3}+ 0.00047203\,{x}^{2}- 0.0030029
\,x \nonumber \\
- 0.30287\,\ln  \left(  593907229\,{x}^{2}- 29656460050\,x
+ 468265974500 \right) 
\nonumber \\
+ 0.19796\,\arctan \left(  0.077829
\,x- 1.9431 \right) + 8.3557
\label{dfastronearly}
\end{eqnarray}
in the nearly Poissonian case  and
\begin{eqnarray}
DF_{non}(x) =
- 0.0000013322\,{x}^{3}+ 0.00033491\,{x}^{2}- 0.014562
\,x
\nonumber \\
- 0.63054\,\ln  \left(  1142509457\,{x}^{2}- 77777418420\,
x+ 2468943718000 \right) 
\nonumber \\
+ 0.52222\,\arctan \left( 
 0.031584\,x- 1.0750 \right) + 18.421
\label{dsastronon}
\end{eqnarray}
in the non Poissonian case.

The random generation of astrophysical chords 
can be implemented by solving in $x$  the following
non linear equation in the nearly Poissonian case
\begin{equation}
R=DF_{nearly}(x)
\quad ,
\label{nlrandomnearly}
\end{equation}
where $R$ is the unit rectangular variate.

The non linear equation to be solved 
in the non Poissonian case is
\begin{equation}
R=DF_{non}(x)
\quad .
\label{nlrandomnon}
\end{equation}

\subsection{The cosmic chord}

In the previous subsection 
we derived a  couple of non linear equations 
for the astrophysical chords, we now outline an 
algorithm for the cosmic environment.
We generated $n$ random chords in Mpc units
solving one of the two non linear equations, 
(\ref{dfastronearly})
or 
(\ref{dsastronon}).
The random chords, denoted by $l_j$, are
\begin{equation} 
l_1 , l_2 \dots l_n
\quad .
\end{equation}
We assume the additivity of the luminosity
distance, $d_{L,j}$, 
and therefore 
\begin{equation}
d_{L,1} = l_1 \quad   d_{L,2} = l_1+l_2 \quad \dots 
d_{L,n} = D_{l,n-1} + l_n
\quad .
\end{equation}

The conversion of the luminosity distance to the redshift 
is made trough formula (\ref{zinverse})
\begin{equation}
z_1=\frac{N(d_{L,1})}{D(d_{L,1})}
\quad
z_2=\frac{N(d_{L,2})}{D(d_{L,2})}
\quad
\dots
\quad
z_n=\frac{N(d_{L,n})}{D(d_{L,n})}
\quad .
\end{equation}

A typical sequence is reported in Table \ref{tablerandom}.

\begin{figure}
\begin{center}
\includegraphics[width=10cm]{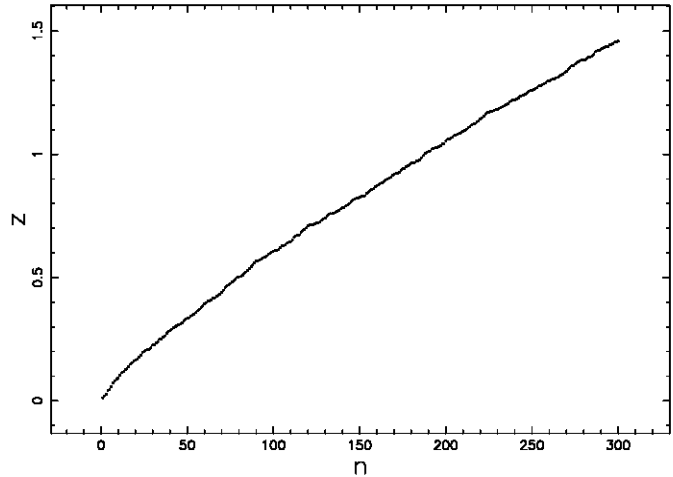}
\end{center}
\caption
{
Value of redshift as function of the number of chords in the 
non Poissonian case.
}
\label{znchords}
\end{figure}

\begin{table}[ht!]
\caption
{
Ten random non Poissonian chords, $l$, 
luminosity distance, $d_L$, and
corresponding redshift in flat cosmology.
}
\label{tablerandom}
\begin{center}
{
\begin{tabular}{|c|c|c|}
\hline
l             & $d_L$ &  z  \\
\hline  
 38.96 &  38.96 & 0.0119 \\
 23.53 &  62.49 & 0.0198 \\
 18.57 &  81.06 & 0.0260 \\
 46.50 & 127.57 & 0.0411 \\
 12.06 & 139.63 & 0.0450 \\
 39.24 & 178.87 & 0.0574 \\
 47.43 & 226.29 & 0.0720 \\
 32.51 & 258.81 & 0.0819 \\
 14.25 & 273.06 & 0.0862 \\
 21.64 & 294.70 & 0.0926 \\
\hline
\end{tabular}
}
\end{center}
\end{table}
A first interesting quantity to be evaluated 
is the number of chords, $N$,
that cover a given interval in redshift,  as an example (0,1),
which is  
\begin{equation}
N = 188 ,\quad non~ Poissonian ~case\quad ;\qquad N = 271,\quad nearly ~Poissonian
~case .    
\end{equation}
Figure \ref{znchords} reports the value reached in redshift 
for rising number of chords.

\subsection{The catalog of galaxies}

\begin{figure*}
\begin{center}
\includegraphics[width=10cm]{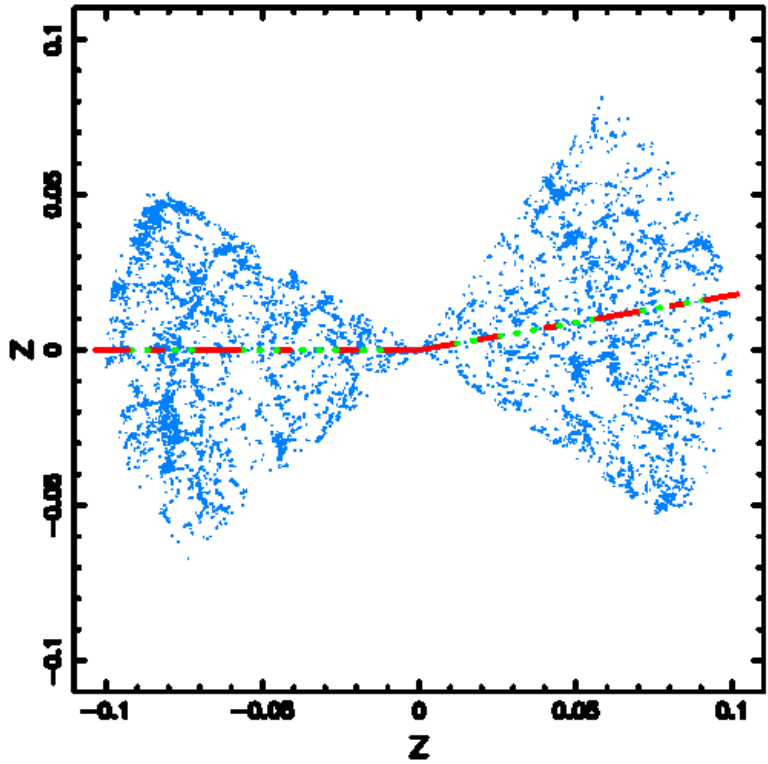}
\end {center}
\caption{
Cone-diagram  of the  galaxies  
in the 2dFGRS  with $z$  $\in$  (0,0.1).  
This plot contains  10234  galaxies (blue color).
Two sequence of chords in the non Poissonian case
are reported, the first chord is red (full line),
the second chord is green            (dotted line). 
}
          \label{corda_2df}%
    \end{figure*}

\begin{figure*}
\begin{center}
\includegraphics[width=10cm]{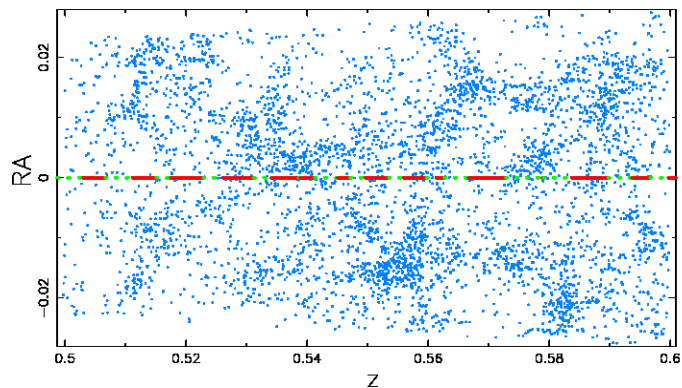}
\end {center}
\caption{
W1 field of VIPERS   with z $\in$  (0.5,0.6).  
This plot contains  4716  galaxies (blue color).
Two sequence of chords in the nearly Poissonian case
are reported, the first chord is red (full line),
the second chord is green (dotted line) . 
}
          \label{corda_vipers}%
    \end{figure*}

A {\it first}  catalog of galaxies is   
the two-degree Field Galaxy Redshift Survey,
in the following 2dFGRS, see \cite{Colless2001},
in the interval $z \in (0.001,0.1)$ :
two strips of  the 2dFGRS are 
shown in Figure~\ref{corda_2df}.
A {\it second}  catalog of galaxies is   
VIMOS Public Extragalactic Survey, in the following VIPERS,
see \cite{Garilli2014}.
We selected the W1 field when
$z$ $\in$  (0.5,0.6),
see Figure \ref{corda_vipers}.

\section{Conclusions}

{\bf Voronoi chords} 
The PDF of the chord length distribution in PVT in the case 
$d=3$,
where $d$ denotes the dimension, is reported as a numerical DF,
see Table 5.7.4  in \cite{Okabe2000}  or as a numerical PDF
see Figure 2 in \cite{Muche2010}.
Here we derived a general expression in the case $d=3$ 
for chord length distribution in the case in which 
the Voronoi volumes are modeled by a Kiang function 
with variable shape parameter $c$.
The generalized PDF is represented by eqn.(\ref{gbscaled}) 
which covers the Poissonian case, $c=5$,
the ordered case    $c>5$ and
the disordered case $c<5$.

{\bf Flat cosmology}
We derived an approximated relationship 
for the luminosity distance 
in spatially flat cosmology with pressureless matter and
the cosmological constant
as function of the redshift, see eqn.(\ref{dlfzminimax})
and the inverse relationship,
the redshift as function of the luminosity distance, 
see eqn.(\ref{zinverse}).

{\bf Astrophysical chord}

In order to find $c$,  the shape parameter of the Kiang's function
for volumes of cosmic voids,  
and $b$, the scale which fix the PDF for astrophysical chord,
we processed two catalogs of cosmic voids,
see Table \ref{bcparameters}.
The exact PDF for astrophysical chord was derived,
see eqns.
(\ref{pdfastrophysicalnearly}) and
(\ref{pdfastrophysicalnon}).
An approximated DF for astrophysical chord  
was deduced in the framework of 
the best minimax rational approximation,
see eqns.
(\ref{dfastronearly})
and
(\ref{dsastronon}).
Two non linear equations, (\ref{nlrandomnon})
and
(\ref{nlrandomnearly}), allow 
the generation of random astrophysical chords.
The inverse formula (\ref{zinverse}) allows the 
superposition of random astrophysical chords
on two catalog of galaxies, 
see Figure~\ref{corda_2df} for 2dFGRS 
and  
Figure~\ref{corda_vipers} for VIPERS.


\end{document}